\documentclass[prd,10pt,twocolumn,nofootinbib,preprint,superscriptaddress]{revtex4}
\pdfoutput=1
\usepackage{feynmp}
\usepackage{amsmath, amssymb, amsthm, graphicx, fancyhdr,epsfig, slashed, mathrsfs}
\usepackage{pifont}
\usepackage{subfigure}
\usepackage{tikzsymbols}
\usepackage{natbib}
\usepackage{float}
\usepackage{lipsum}
\usepackage{mathtools} 
\usepackage{setspace}
\usepackage[normalem]{ulem}
\usepackage{tabularx}
\usepackage{tikz,xcolor,hyperref}
\usepackage{cancel}
\hypersetup{
     colorlinks=true,
     citecolor    = blue,
     linkcolor= blue
}

\definecolor{lime}{HTML}{A6CE39}
\DeclareRobustCommand{\orcidicon}{
	\begin{tikzpicture}
	\draw[lime, fill=lime] (0,0) 
	circle [radius=0.2] 
	node[white] {{\fontfamily{qag}\selectfont \tiny ID}};
	\draw[white, fill=white] (-0.0625,0.095) 
	circle [radius=0.007];
	\end{tikzpicture}
	\hspace{-2mm}
}

\foreach \x in {A, ..., Z}{\expandafter\xdef\csname orcid\x\endcsname{\noexpand\href{https://orcid.org/\csname orcidauthor\x\endcsname}
			{\noexpand\orcidicon}}
}

\begin{document}

\title{Cosmic-ray-electron boosted light dark matter: Implications of LZ 2025 data}

\author{Sk Jeesun\orcidA}
\email{jeesun@sjtu.edu.cn}
\affiliation{State Key Laboratory of Dark Matter Physics, Tsung-Dao Lee Institute $\&$ School of Physics and Astronomy, Shanghai Jiao Tong University, Shanghai 200240, China}
\affiliation{Key Laboratory for Particle Astrophysics and Cosmology (MOE) $\&$ Shanghai Key Laboratory for Particle Physics and Cosmology, Shanghai Jiao Tong University, Shanghai 200240, China}
\author{Anirban Majumdar\orcidB}
\email{anirban19@iiserb.ac.in}
\affiliation{Department of Physics, Indian Institute of Science Education and Research - Bhopal,
Bhopal Bypass Road, Bhauri, Bhopal 462066, India}

\begin{abstract}
    Current multiton detectors put stringent constraints on the GeV-scale galactic dark matter, pushing the allowed cross section almost toward the neutrino fog, yet remain mostly insensitive to the light dark matter. 
    Cosmic rays can upscatter the nonrelativistic halo dark matter particles, making a subpopulation of them gain sufficient kinetic energy to be discernible in current
 direct search experiments. 
 In this work, we explore this alternate strategy to probe sub-MeV electrophilic dark matter boosted by cosmic rays with the latest data of LZ 2025 (WS2024 run).
     We also incorporate the attenuation effect on the boosted dark matter flux during its propagation through the Earth and perform a full numerical treatment to obtain the resulting event rate. Our result shows LZ 2025 data improve the constraint on the MeV scale dark matter by almost $\sim\mathcal{O}(1)$ compared to the previous XENONnT limit for the energy-independent cross section.
    Using realistic energy-dependent cross sections, we also analyze such a scenario, where the associated mediator mass plays a crucial role in governing the event rate and hence the expected limits too.
    With energy-dependent cross sections, our obtained limits also remain stronger than the existing constraints from the XENONnT experiment.
    Even compared to the limits from neutrino detectors with much larger target masses, LZ 2025 can place stringent constraints in certain regions of the mediator parameter space, particularly in the light-mediator regime, excluding previously unexplored regions.
\end{abstract}

\maketitle

{\bf Introduction:}
The quest for the particle nature of the nonluminous and nonbaryonic dark matter (DM) persists even after the numerous investigations over the past several decades \cite{Jungman:1995df,Cirelli:2024ssz}.
DM constitutes $25\%$ of the total energy budget of the Universe at present and has been strongly indicated by observations relying on its gravitational interactions \cite{Zwicky:1933gu,Rubin:1970zza,PAMELA:2011bbe,Planck:2018vyg}.
Despite the strong observational evidence, the nongravitational interactions of particle DM still remain elusive.
Ground-based direct detection (DD) experiments bring an excellent opportunity to trace such galactic DM particles through their scatterings with nucleons or electrons while the Earth traverses through the DM halo \cite{Cirelli:2024ssz,Lin:2019uvt,Feng:2010gw, AristizabalSierra:2021kht, DeRomeri:2025nkx}.
Null results in experiments like XENONnT \cite{XENON:2020iwh}, LUX-ZEPLIN (LZ) \cite{LZ:2022lsv,LZ:2025zpw} ,
PandaX-4T \cite{PandaX:2025rrz}, and DarkSide-50 \cite{DarkSide:2018ppu} have placed stringent constraints on the electroweak scale DM and thus motivate the community to seek DM particles in the sub-GeV mass range.

The non relativistic galactic halo DM particles with MeV scale mass and velocity $\sim 10^{-3}c$  fail to generate a recoil energy above the threshold energy ($\gtrsim$ keV) of the multiton detectors making the existing direct searches ineffective in detecting light halo DM particles.
To circumvent this deadlock, a subdominant boosted component of light DM with almost relativistic speeds is being explored \cite{Bringmann:2018cvk,Ema:2018bih}.
MeV-scale DM particles in the halo with electronic or nuclear interactions are expected to inevitably scatter with highly energetic (with energy $\gtrsim$  GeV-TeV) cosmic rays in the Galaxy and thus may gain kinetic energies higher than the experimental thresholds.
This subpopulation of light DM reaching the Earth-based multiton detectors and
producing a unique recoil signature may help to resolve the intriguing puzzle of the light DM. 
Null observations of any such recoil lead to stringent constraint on the respective DM interactions \cite{Bringmann:2018cvk, Ema:2018bih,Dent:2019krz,Dent:2020syp,Bardhan:2022bdg,Bell:2020rkw,Ghosh:2024dqw}.
Numerous works in the recent past have explored the novel idea of DM boosted by other mechanisms as well, to probe such light DM interactions. Examples include DM boosted by blazar jet \cite{Wang:2021jic,Herrera:2023nww,Jeesun:2025gzt,DeMarchi:2025uoo}, diffuse supernova neutrinos \cite{Das:2021lcr, DeRomeri:2023ytt}, supernova emission \cite{Bhalla:2025vnq} , and solar reflection \cite{An:2017ojc,Emken:2024nox}.

In this work, we explore an electrophilic light DM boosted by the cosmic ray electrons (CRe) and their possible signature in the LZ experiment \cite{LZ:2025zpw}.
LZ, with its recent data from WS2024, which has a 3.3 ton-year exposure, has the potential to probe such low-mass DM parameter space, which serves as the main goal of this work. 
Existing literature has already shown that the DM recoil signature as well as the obtained limits, change drastically in the presence of a realistic energy-dependent cross section compared to a constant cross section \cite{Dent:2020syp,Ghosh:2021vkt,Das:2024ghw}.
Apart from a constant cross section, we  consider two realistic cross sections between DM and electron (e): (1) vector mediator and (2) scalar mediator.
We emphasize that in such boosted DM (BDM) scenarios, the mass of the associated mediator particle plays a nontrivial role in governing the recoil rate and, consequently, in the obtained constraints.
To give a rigorous example, we present a benchmark vector mediator model and show the constraints in the mediator parameter space.
We also incorporate the Earth's attenuation effect on the boosted dark matter flux during its propagation and perform a full numerical treatment to obtain the limits.
Though multiton neutrino detectors like Super Kamiokande and IceCube provide the strongest constraint on such BDM, we show that in certain parameter space, LZ 2025 \cite{LZ:2025zpw}, with its low threshold energy and sizable exposure, places the improved direct detection constraint on MeV-scale electrophilic DM.

{\bf Cosmic ray boosted DM flux:} A subdominant yet energetic component of the cosmic ray is the CRe as inferred from observation \cite{Boschini:2017fxq}. DM having an interaction channel with electrons is expected to be upscattered by this CRe, leading to a boosted flux of $\chi$ given as \cite{Bringmann:2018cvk},
\begin{equation}
    \frac{d\Phi_\chi}{dT_\chi}= D_{\rm eff} \frac{\rho_\chi^{\rm local}}{m_\chi} \int^{\infty}_{T_e^{\rm min}(T_\chi)} \frac{d\Phi_e^{\rm CR}}{dT_e} \frac{d\sigma_{\chi e}}{dT_\chi} dT_e,
    \label{eq:dm_flux}
\end{equation} 
where $\rho_\chi^{\rm local}=0.3~{\rm GeV/cm}^3$ is the local DM density whereas, $m_\chi$ and $\sigma_{\chi e}$ signify  the DM mass and elastic DM-$e$ scattering cross section respectively.
$T_\chi$ and $T_e$  represent the kinetic energy of the boosted (upscattered) DM and CRe, respectively. 

$D_{\rm eff}$ is the effective  diffusion zone parameter given by the line of sight (LOS) integral of the Galactic DM density profile \cite{Bringmann:2018cvk,Dent:2020syp},
\begin{equation}
\label{eq:Dfactor}
D_{\rm eff}
= \frac{1}{\rho_\chi^{\rm local}}\int \frac{d\Omega}{4\pi}
\int_0^{\ell_{\rm max}} \rho_{\rm MW}\,d\ell\,.
\end{equation}
$\rho_{\rm MW}$ is the DM density distribution in the Milky Way and we consider the conventional Navarro-Frenk-White (NFW) DM profile which is a function of the galactocentric distance along the LOS $r$ \cite{Navarro:1996gj}. 
$r$ is given as $
r(\ell,\psi)
= \sqrt{r_\odot^2 - 2\ell r_\odot \cos\psi + \ell^2}\,$, where $r_\odot = 8.5~\mathrm{kpc}$ and $\psi$ represent the Sun’s distance from the Galactic center and the observation angle, respectively.
$D_{\rm eff}$ in principle incorporates the effective distance from Earth, up to which the effects from the CRe-DM scattering are considered. 
With a NFW profile and integrating over the LOS integral up to $1$ kpc leads to $D_{\rm eff}\sim 1$ kpc.
The same with an optimistic upper limit up to 10 kpc leads to $D_{\rm eff} \sim 10$ kpc.
For this work, we simply consider a homogeneous distribution of the CRe flux as inferred from the local observations, with $D_{\rm eff}= 1$ kpc for the conservative limit \cite{Bringmann:2018cvk,Dent:2020syp}.

\begin{figure}[H]
    \centering
    \includegraphics[scale=0.45]{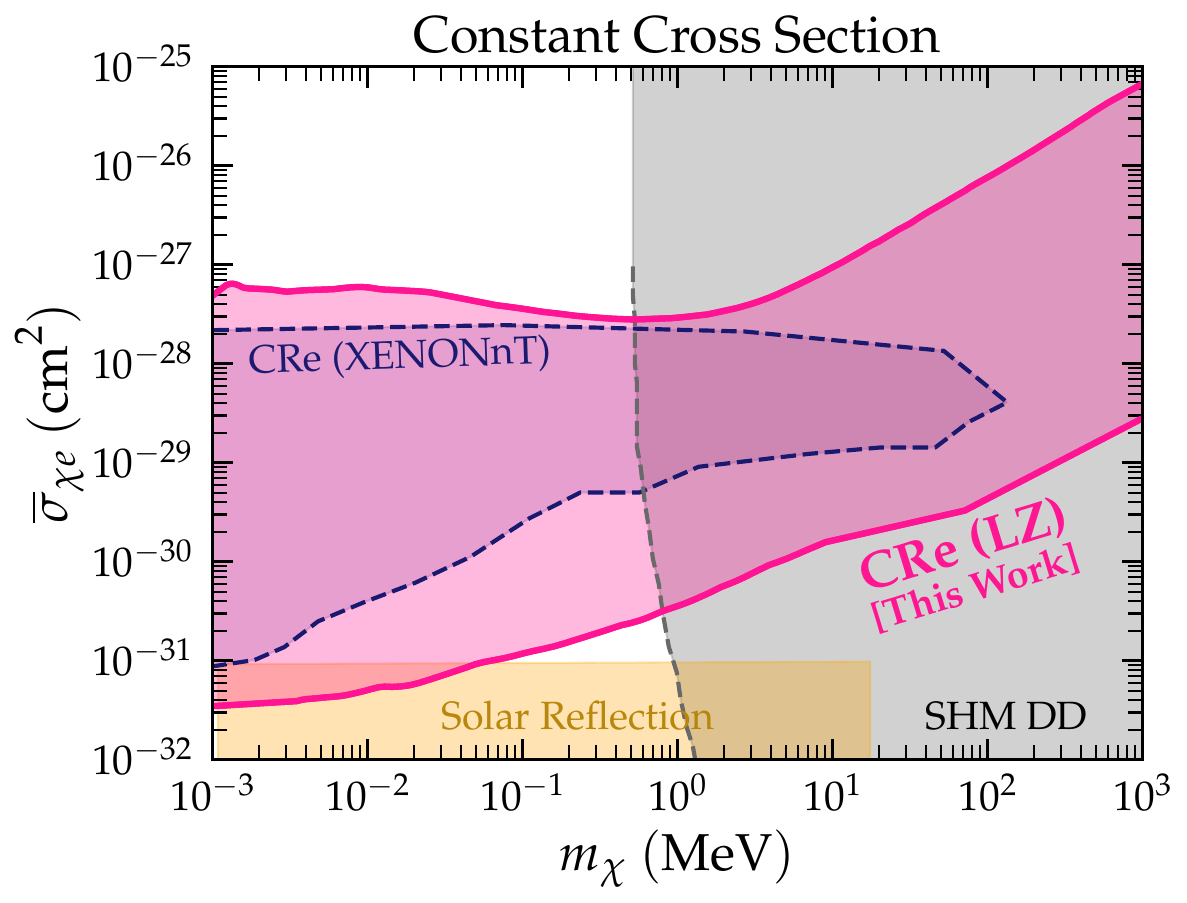}
    \caption{$2\sigma$ C.L. constraints on CRe boosted DM with {\bf constant} cross section from LZ 2025 is shown by the magenta shaded region. Existing constraints from the BDM search at XENONnT \cite{Guha:2024mjr} (dashed blue line), solar reflected DM \cite{An:2017ojc} (orange shaded region) and the combined
DD limits from SuperCDMS \cite{SuperCDMS:2018mne}, SENSEI \cite{Crisler:2018gci}, and XENON10 \cite{Essig:2012yx} for Galactic halo DM (gray shaded region) are also shown.}
    \label{fig:LZ_con}
\end{figure}

The term $d\Phi_e^{\rm CR}/dT_e$ in  Eq.~\eqref{eq:dm_flux} signifies the local interstellar flux of cosmic electrons and is obtained from the observations in experiments like Voyager \cite{Cummings:2016pdr}, Fermi-Lat \cite{Fermi-LAT:2011baq}, PAMELA \cite{PAMELA:2011bbe}, and AMS-02 \cite{AMS:2014gdf}.
Fitting the observational data the CRe flux is obtained as \cite{Boschini:2017fxq, Boschini:2018zdv,Bardhan:2022bdg},
\begin{widetext}
\begin{equation}
    \dfrac{d\Phi_e^{\rm CR} }{dT_e d\Omega} = 
    \begin{cases}
    \dfrac{1.799\times 10^{44} T_e^{-12.061}}{\big(1+2.762\times 10^{36} T_e^{-9.269}
    +3.853\times 10^{40} T_e^{-10.697}\big)} &~{\rm for~ } T_e<6880 {\rm ~MeV}\\
    3.259 \times10^{10} T_e^{-3.505} +3.204\times10^5 T_e^{-2.620} &~{\rm for~ } T_e \geq 6880 {\rm ~MeV} 
    \end{cases}
\end{equation}
\end{widetext}
in units of $ {\rm sr}^{-1} {\rm m}^{-2} {\rm s}^{-1} {\rm MeV}^{-1}$, where $T_e$ is given in MeV.
As mentioned earlier, we treat this flux to be homogeneous with $D_{\rm eff}=1$ kpc.

The minimum kinetic energy of CRe required to boost a non relativistic DM to energy $T_\chi$ is denoted as  $T_e^{\rm min}$ in Eq.~\eqref{eq:dm_flux} and is given as \cite{Bringmann:2018cvk},
\begin{equation}
    T_{e}^{\rm min}(T_\chi) =
    \left(\frac{T_\chi}{2}-m_e \right) \left[1 \pm \sqrt{1+\frac{2 T_\chi (m_e+m_\chi)^2}{m_\chi(2 m_e-T_\chi)^2}}~~\right] ,
    \label{eq:tmin}
\end{equation}
The $+$ve ($-$ve) sign in the above equation applies when $T_\chi > 2m_e$ ($T_\chi < 2m_e$).

The BDM flux in Eq.~\eqref{eq:dm_flux} has a crucial dependence also on the differential cross section between $e$ and $\chi$
where the underlying Lorentz structures of the BSM interaction between DM and $e$ come into play.
In accordance with the GeV scale non relativistic (NR) DM scattering, usually the energy-independent differential cross section between CRe and $\chi$ is considered in the literature which reads as ~\cite{Bringmann:2018cvk}, 

\begin{equation}
    \frac{d\sigma_{\chi e}}{dT_\chi}=\frac{\bar{\sigma}_{\chi e}}{T_\chi^{\rm max}}.
    \label{eq:cons}
\end{equation}
$T_\chi^{\rm max}$ is the maximum kinetic energy $\chi$ can achieve after the scattering, given as
\begin{eqnarray}
~T_\chi^{\rm max} =\dfrac{T_e^2+2 m_e T_e}{T_e+(m_e+m_\chi)^2/(2 m_\chi)}.
\end{eqnarray}

Since CRe has sufficient energy to boost a light (sub-GeV) DM to gain almost semirelativistic velocities, the differential cross sections may differ significantly in cases where kinetic energy of the incoming (or, the target) particle appears in the amplitude  often referred as 
energy-dependent cross sections \cite{Dent:2019krz,Bardhan:2022bdg}. 
This often leads to significant alteration in the obtained limits, especially in the low DM mass region, compelling us to consider  
the energy-dependent cross sections as well.
In this work, we consider the scalar and vector mediated DM-e scattering whose differential cross sections read as,
\begin{widetext}
    \begin{align}
    \left(\frac{d\sigma_{\chi e}}{dT_\chi}\right)_{\rm scalar} &= \bar{\sigma}_{\chi e} \left(\dfrac{m_\chi}{4\mu_{e \chi}^2}\right) \dfrac{(2m_\chi+T_\chi)(2 m_e^2+m_\chi T_\chi)}{s_{\rm CR}T_\chi^{\rm max}}~F^2_{\rm DM}(q^2) \nonumber\\ 
    \left(\frac{d\sigma_{\chi e}}{dT_\chi}\right)_{\rm vector} &= \bar{\sigma}_{\chi e} \left(\dfrac{m_\chi}{2\mu_{e \chi}^2}\right)  \dfrac{ (2 m_\chi (m_e+T_e)^2 -T_\chi ((m_e+m_\chi)^2+2 m_\chi T_e)+m_\chi T_\chi^2 )}{s_{\rm CR} T_\chi^{\rm max} }~F^2_{\rm DM}(q^2),
    \label{eq:deps}
\end{align}
\end{widetext}
respectively, with $\mu_{e\chi}$ being the reduced mass of the $\chi-e$ system. $s_{\rm CR}=(m_\chi+m_e)^2+2 m_\chi T_e$ is the center of momentum energy of $\chi-e$ scattering.
$F_{\rm DM}(q^2)$ incorporates the information of the mass scale of the mediator participating in the aforementioned interaction, given as,
\begin{eqnarray}
    F_{\rm DM}(q^2)=\frac{q_{\rm ref}^2+m_{\rm med}^2}{q^2+m_{\rm med}^2},
\end{eqnarray}
with $q=\sqrt{2 m_\chi T_\chi}$ being the associated three-momentum transfer. $q_{\rm ref}=\alpha m_e$ is the reference momentum to define the NR cross section $\bar{\sigma}_{\chi e}$.
$m_{\rm med}$ is the mediator mass.
Note that for $m_{\rm med}\gg q$ ($m_{\rm med}\ll q_{\rm ref}$), i.e. in the heavy (light) mediator limit $F_{\rm DM}(q^2)$ reduces to 1 ($(\alpha m_e/q)^2$).

{\bf Attenuation:}
The incoming BDM flux with high cross section with matter is expected to be affected by the attenuation effect of the Earth's material while traversing through it before reaching the underground detector\footnote{In principle, the incoming BDM flux may also undergo attenuation while traversing the Earth's atmosphere. However, owing to the significantly lower number density of atmospheric matter compared to that of the Earth~\cite{DeRomeri:2023ytt}, the corresponding attenuation is negligible for the range of cross sections considered in this work. Therefore, we neglect atmospheric attenuation throughout our analysis.
}.
This leads to loss of kinetic energy (KE) of the incoming $\chi$. The KE of the BDM particles after traveling a distance $z$ within the Earth is governed by the 
energy loss equation\footnote{Since we consider an electrophilic dark matter scenario, the attenuation is evaluated through DM-electron elastic scatterings inside the Earth. Possible contributions from DM-nucleus interactions are therefore not applicable in the present setup.
} \cite{Bringmann:2018cvk,Xia:2021vbz}, 
\begin{eqnarray}
    \frac{d T_\chi^z}{d z} = - n_e \int_{0}^{T_{er}^{\rm max}(T_\chi^z)} \frac{d \sigma_{\chi e}}{d T_{er}}~ T_{er}~ dT_{er},
    \label{eq:attn}
\end{eqnarray}
with $z$ given as \cite{DeRomeri:2023ytt},
\begin{eqnarray}
    z=-(R_{\oplus}-h_d)\cos{\theta_z}+\sqrt{R_{\oplus}^2-(R_\oplus-h_d)^2\sin^2{\theta_z}}.
    \nonumber
\end{eqnarray}
Here, $T_{er}$ signifies the recoil energy of the outgoing electrons after being scattered by $\chi$ within the Earth, while $T_{er}^{\rm max}(T_\chi^z)$ corresponds to the maximum recoil energy of the outgoing electrons and can be obtained by,
\begin{eqnarray}
~T_{er}^{\rm max}(T_\chi) =\dfrac{T_\chi^2+2 m_\chi T_\chi}{T_\chi+(m_e+m_\chi)^2/(2 m_e)}.
\end{eqnarray}
$d\sigma_{\chi e}/dT_{er}$ represents the differential cross section of energetic incoming DM with the electron. Here also, the differential scattering cross section between the incoming energetic DM and the target electron at rest, plays the crucial role in the event rates.
For energy independent amplitude the differential cross section is given as \cite{Bringmann:2018cvk},
\begin{equation}
    \frac{d\sigma_{\chi e}}{dT_{er}}=\frac{\bar{\sigma}_{\chi e}}{T_{er}^{\rm max}}.
    \label{eq:cons}
\end{equation}
Energy dependent cross sections with target electrons for scalar and vector mediators can be obtained from earlier discussed Eq.~\eqref{eq:deps} with the substitution $T_\chi\to T_{er},~T_e\to T_\chi,~m_e\to m_\chi,~ m_\chi\to m_e,~ T_\chi^{\rm max}\to T_{er}^{\rm max}$ and $s_{\rm CR} \to s_{\chi}=(m_\chi+m_e)^2+2 m_e T_\chi$ [see Eq.~\eqref{eq:sig_scalar} and Eq.~\eqref{eq:sig_vector}].
Finally, the average electron density within the Earth is considered as $n_e(=8\times 10^{23}~{\rm cm}^{-3}$) ~\cite{DeRomeri:2023ytt}.
In the expression of $z$, the parameters like $\theta_z$, $R_{\oplus}$, and $h_d$ signify the zenith angle of the incoming BDM particles, Earth's radius, and the depth of the detector at zero zenith angle, respectively.
For LZ, we consider $h_d = 1.47~\mathrm{km}$. 

The energy loss equation [Eq.~\eqref{eq:attn}] is solved with the initial condition, $T_\chi^z(z=0)\equiv T_\chi^0$, which describes the KE of $\chi$ at Earth's surface.
To obtain the realistic recoil rates, one needs to obtain the differential flux after attenuation ~\cite{Bringmann:2018cvk,DeRomeri:2023ytt},
\begin{equation}
\frac{d \Phi_\chi}{d T_\chi^z}=\int d \Omega \left.\frac{d ^2\Phi_\chi}{d T_\chi d\Omega}\right|_{T_\chi^0} \frac{d T_\chi^0}{d T_\chi^z}.
\label{eqn:Flux_att}
\end{equation}
Note that, $\left.\frac{d ^2\Phi_\chi}{d T_\chi d\Omega}\right|_{T_\chi^0}$ stands as the usual incoming flux of CRe boosted $\chi$ with KE ${T_\chi^0}$ described in
Eq.~\eqref{eq:dm_flux}.
In the limiting case, $(m_\chi+m_e)^2\ll 2 m_e T_\chi^z$ and for energy-independent cross section approximation the energy loss equation [Eq.~\eqref{eq:attn}] can be solved analytically, leading to a simplified expression of DM flux \cite{Bringmann:2018cvk,Ghosh:2021vkt, DeRomeri:2023ytt}. However, such an approximation is not guaranteed in the whole parameter space, especially for energy-dependent interactions.
Hence, we refrain from using such analytical formalism of attenuation effect in this work.

To obtain the numerical solution and perform the full parameter-space scan, we proceed as follows. For a fixed $m_\chi$ and $\bar{\sigma}_{\chi e}$, we numerically solve Eq.~\eqref{eq:attn} over a grid of initial kinetic energies $T_\chi^0\in[10~{\rm eV},10^5~{\rm GeV}]$ and propagation distances $z$, obtaining $T_\chi^z$ as a function of $(z,T_\chi^0)$. The grid is then inverted to interpolate $T_\chi^0(z,T_\chi^z)$ and compute the Jacobian $d T_\chi^0/d T_\chi^z$, which is used in Eq.~\eqref{eqn:Flux_att} to derive the attenuated underground flux after zenith-angle integration. 
The variation in attenuated BDM flux for different interactions can be found in Appendix \ref{sec:attenuation}.
Finally, using the attenuated flux, we simulate the recoil spectrum as discussed in the following [Eq.~\eqref{eq:recoil}] and perform the numerical scan to extract the corresponding constraints \footnote{
The detailed methodology can be found in Ref.\cite{Jeesun:2026wcx}.}.

{\bf Recoil spectra:} Upon arriving at the Earth-based underground DM detector such BDM particles will scatter with the target electrons in the detector and produce recoil signature.
The expected event rate is calculated as,
\begin{equation}
    \frac{d R_{\chi e}}{d T_{er}}=t_{\rm exp} n_{\rm T} Z_{\rm eff}(T_{er}) \mathcal{E}(T_{er}) \int^{T_{\chi}^{\infty}}_{T_{\chi}^{\rm min}(T_{er})}d T_{\chi}\, \frac{d\Phi_\chi}{d T_\chi^z} \frac{d \sigma_{\chi e}}{d T_{er}}\,,
    \label{eq:recoil}
\end{equation}
where the true recoil energy of the target electrons, the experimental run time and the number of  target atoms in the detector fiducial volume are represented as  $T_{er}$, $t_{\rm exp}$ and $n_{\rm T}$  respectively, while  the binding energy effects of atomic electrons in xenon are incorporated through the effective charge function $Z_{\rm eff}(T_{er})$. Following Ref.~\cite{Chen:2016eab}, this function is approximated using a series of step functions that account for the ionization thresholds of individual atomic shells, and is expressed as $Z_{\rm eff}(T_{er}) \approx \sum_{i=1}^{54} \Theta\!\left(T_{er}-\mathscr{B}_i\right)$, where $\Theta(x)$ denotes the Heaviside step function. The quantities $\mathscr{B}_i$ correspond to the single-particle binding energies of the $i$th electron in the xenon atom, evaluated using Hartree-Fock atomic structure calculations as reported in Ref.~\cite{Chen:2016eab}. $\mathcal{E}(T_{er})$ encodes the detector efficiency as a function of recoil energy. In this work, $\mathcal{E}(T_{er})$ is taken from the publicly reported 1D region-of-interest (ROI) efficiency curve shown in Fig.~1 of the LZ WS2024 data release~\cite{LZ:2025zpw}, which incorporates the combined effects of trigger response, S1 threshold requirements, single-scatter reconstruction, and data analysis selection cuts.
$T_{\chi}^{\rm min}$ stands for the  minimum kinetic energy of $\chi$ needed to scatter an electron with a recoil energy $T_{er}$ and can be found from Eq.~\eqref{eq:tmin} with the substitution $T_\chi\to T_{er},~m_e\to m_\chi$ and $m_\chi\to m_e$
\cite{Bringmann:2018cvk}.

To simulate the events, we use the updated WS2024 data from the LZ Collaboration corresponding to their 1D ROI likelihood analysis in reconstructed energy, published in 2025 with a total exposure of $3.3~\mathrm{ton}\times\mathrm{year}$~\cite{LZ:2025zpw}.
To obtain the reconstructed event spectra we use the smearing,
\begin{eqnarray}
    \frac{dR_{\chi e}}{dT_{er}^{\rm reco}} =\int_{0}^{T_{er}^{\rm max}} dT_{er} ~\frac{dR_{\chi e}}{dT_{er}} \mathcal{G}(T_{er},T_{er}^{\rm reco}),
    \label{eq:smear}
\end{eqnarray}
$T_{er}^{\rm reco}$ and $\mathcal{G}(T_{er},T_{er}^{\rm reco})$ are the reconstructed electron recoil energy and the Gaussian smearing function respectively. 
%
We consider the resolution power $\sigma=0.323\sqrt{T_{er}^{\rm reco}}$ keV~\cite{Pereira:2023rte}.
Using the BDM flux of Eq.~\eqref{eq:dm_flux} and incorporating the Earth attenuation effect described in Eq.~\eqref{eqn:Flux_att}, we evaluate the recoil spectrum through Eq.~\eqref{eq:recoil}. Furthermore, after applying the detector smearing as in Eq.~\eqref{eq:smear}, we obtain the expected number of events in each bin for a fixed choice of ($m_\chi$) and ($\bar{\sigma}_{\chi e}$).
To place the constraints, we compute the $\chi^2$ value at each point in the $m_\chi$--$\bar{\sigma}_{\chi e}$ plane and marginalize over the nuisance parameters associated with the uncertainties of the various background components. We then compute $\Delta\chi^2=\chi^2-\chi^2_{\rm min}$, where $\chi^2_{\rm min}$ denotes the global minimum of the $\chi^2$ function over the entire parameter space. The $2\sigma$ confidence-level (C.L.) exclusion contour is obtained by selecting the parameter points satisfying $\Delta\chi^2=6.18$.
A detailed discussion about the recoil spectra and $\chi^2$ analysis can be found in Appendix \ref{sec:recoil},
and Appendix \ref{sec:chi} respectively.

{\bf Constraints:} 
We show our limits obtained at $2\sigma$ C.L. from LZ 2025 in Fig.~\ref{fig:LZ_con}, considering a constant cross section depicted by the magenta shaded region. We also showcase the existing limit on CRe boosted DM from XENONnT \cite{Guha:2024mjr} shown by dashed blue lines. Our obtained limit excludes $\sim\mathcal{O}(1)$ smaller $\bar{\sigma}_{\chi e}$ than the XENONnT for $m_\chi \sim~0.1$ MeV. 
With a decrease in $m_\chi$ the bound 
on $\bar{\sigma}_{\chi e}$ becomes stronger because of the higher amount of incoming BDM flux for light $\chi$ \cite{Bardhan:2022bdg}.
Note that, for a higher value of the cross section (e.g. $\bar{\sigma}_{\chi e} \gtrsim 10^{-27}{\rm cm}^2$ for $m_\chi=1$ MeV) the incoming DM flux gets significantly attenuated, failing to generate any statistically significant event at the detector. Hence, apart from the upper limit on $\bar{\sigma}_{\chi e}$, we also observe an attenuation ceiling in the aforementioned plot.
Complementary bounds from solar reflected DM \cite{Emken:2024nox} (orange shaded region) and the combined
DD limits from SuperCDMS \cite{SuperCDMS:2018mne}, SENSEI \cite{Crisler:2018gci}, and XENON10 \cite{Essig:2012yx} for Galactic halo DM (gray shaded region) are also portrayed for comparison.

\begin{figure*}[htp]
    \centering
    \subfigure[]{
    \includegraphics[scale=0.43]{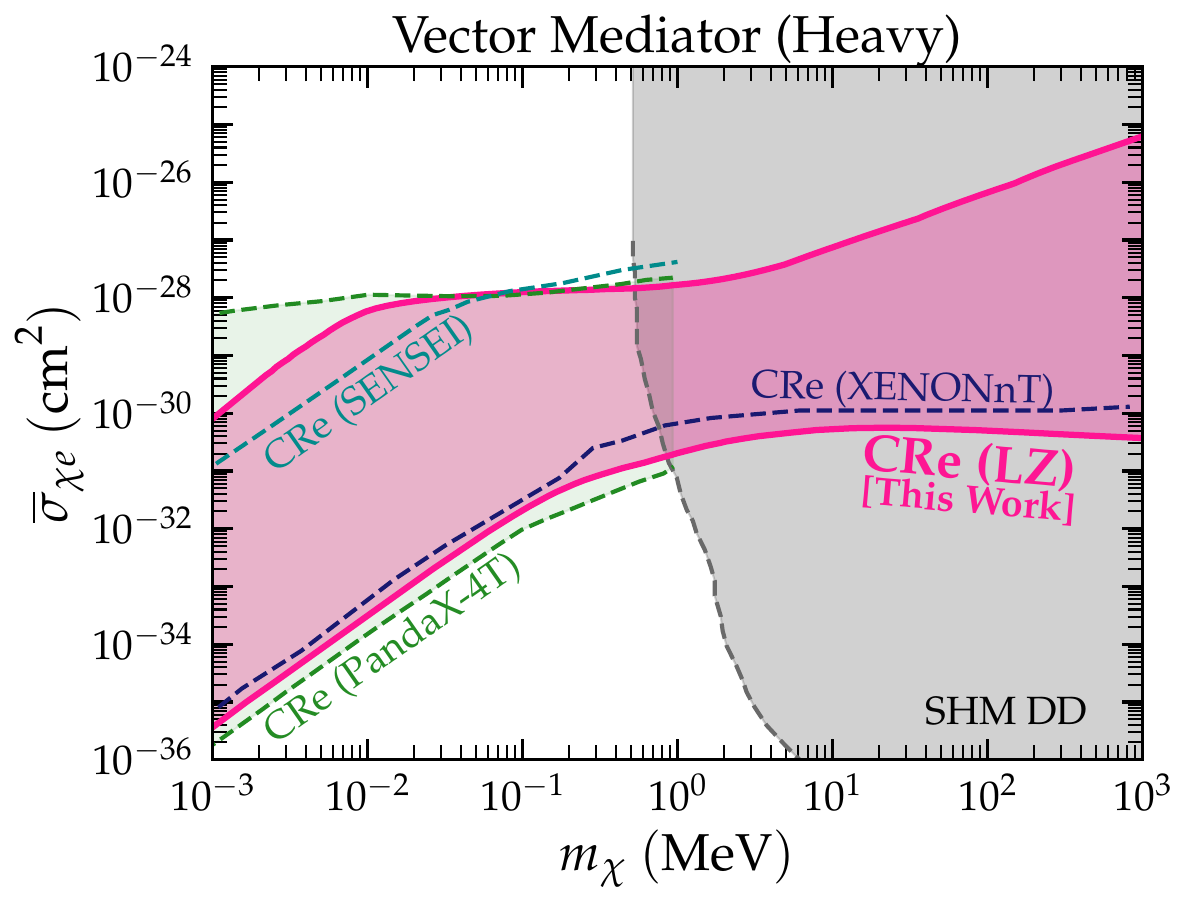}}
    \subfigure[]{
    \includegraphics[scale=0.43]{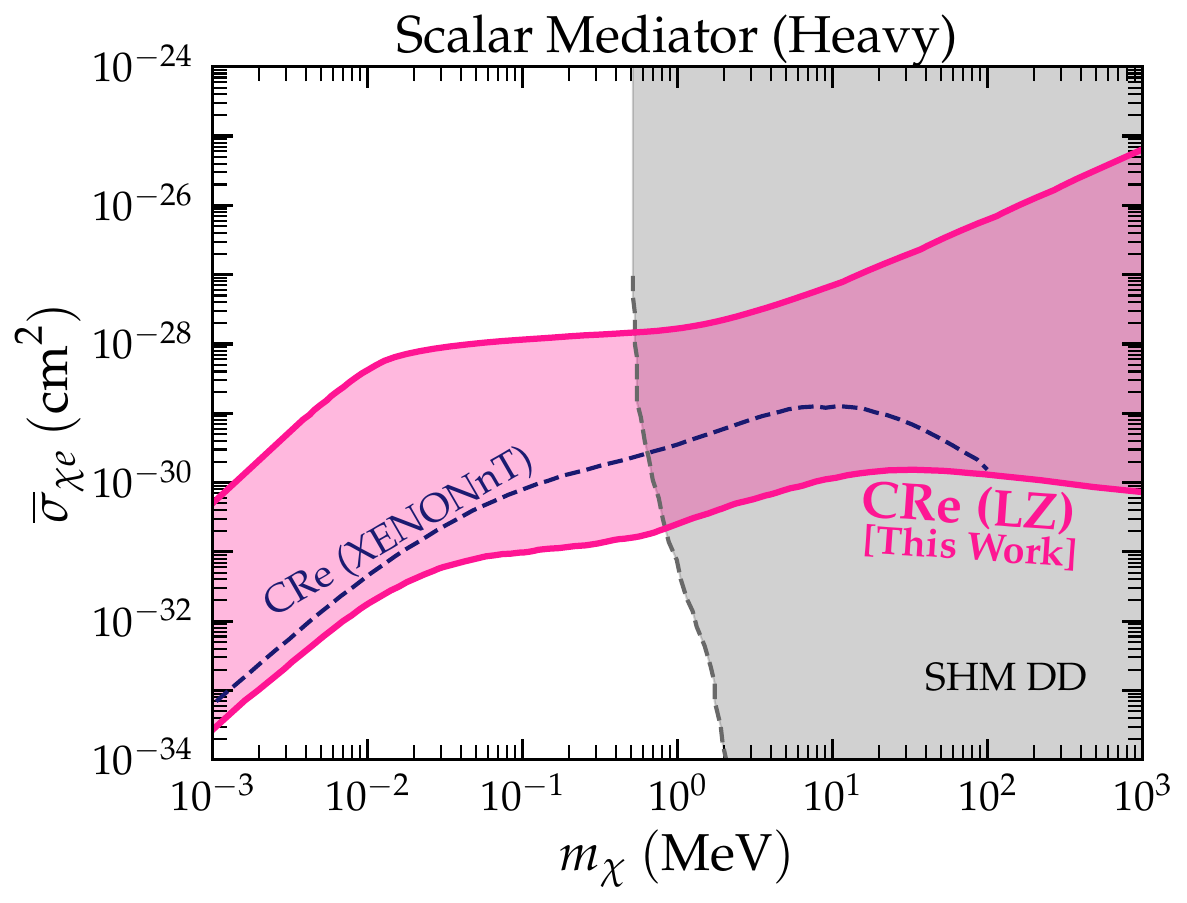}}
    \caption{{$2\sigma$ C.L. constraints on CRe boosted DM in the heavy mediator limit ($F_{\rm DM}(q)=1$) with (a) {\bf vector} mediator and (b)  {\bf scalar} mediator from LZ 2025 are shown by the magenta shaded region. Existing constraints from the BDM search at XENONnT \cite{Guha:2024mjr} (dashed blue line); PandaX-4T \cite{PandaX:2024pme} (green dashed lines); utilizing plasmon enhancement in SENSEI \cite{Liang:2024xcx}(dark cyan dashed lines); and the combined halo
DD limits from SuperCDMS \cite{SuperCDMS:2018mne}, SENSEI \cite{Crisler:2018gci}, and XENON10 \cite{Essig:2012yx} for Galactic halo DM (gray shaded region) are also portrayed.}}
    \label{fig:LZ_heavy}
\end{figure*}
\begin{figure*}[htp]
    \centering
    \subfigure[]{
    \includegraphics[scale=0.43]{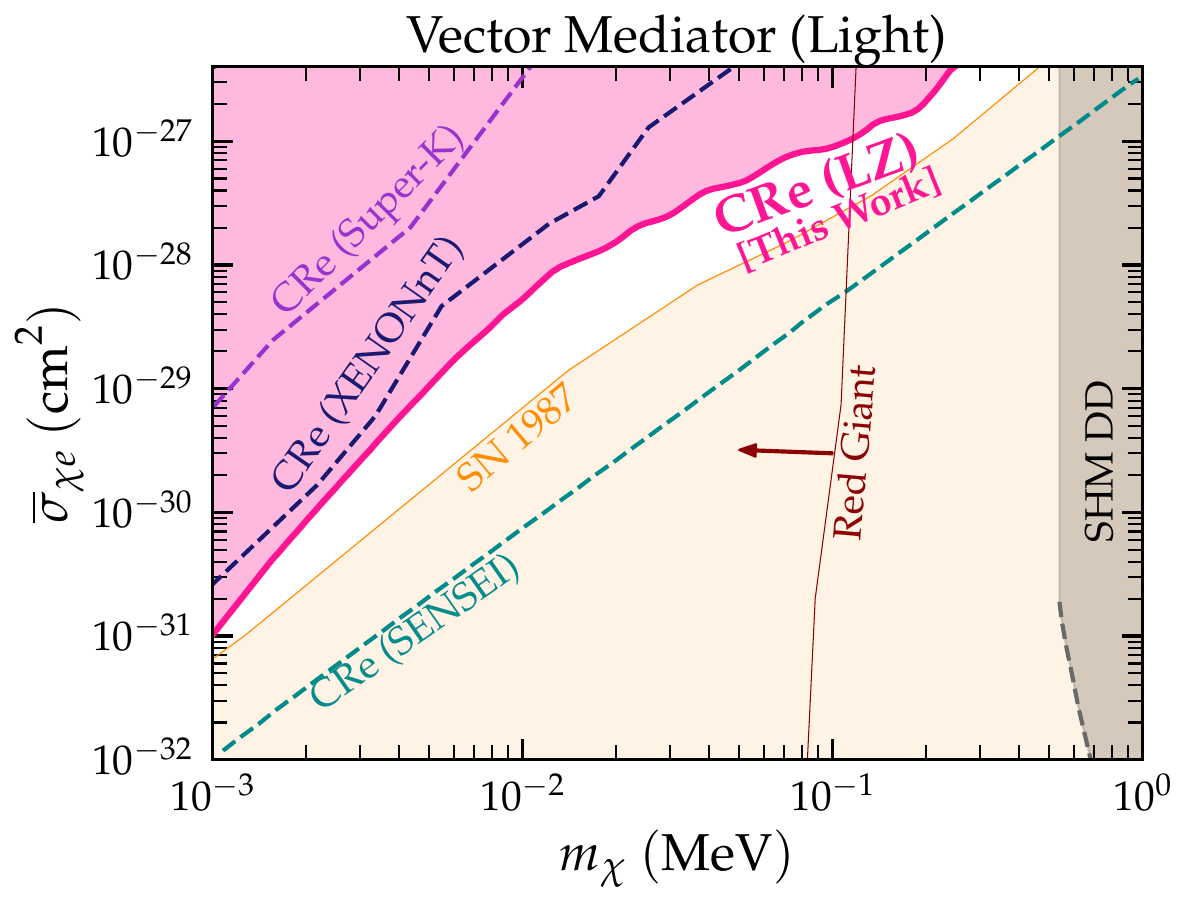}}
    \subfigure[]{
    \includegraphics[scale=0.43]{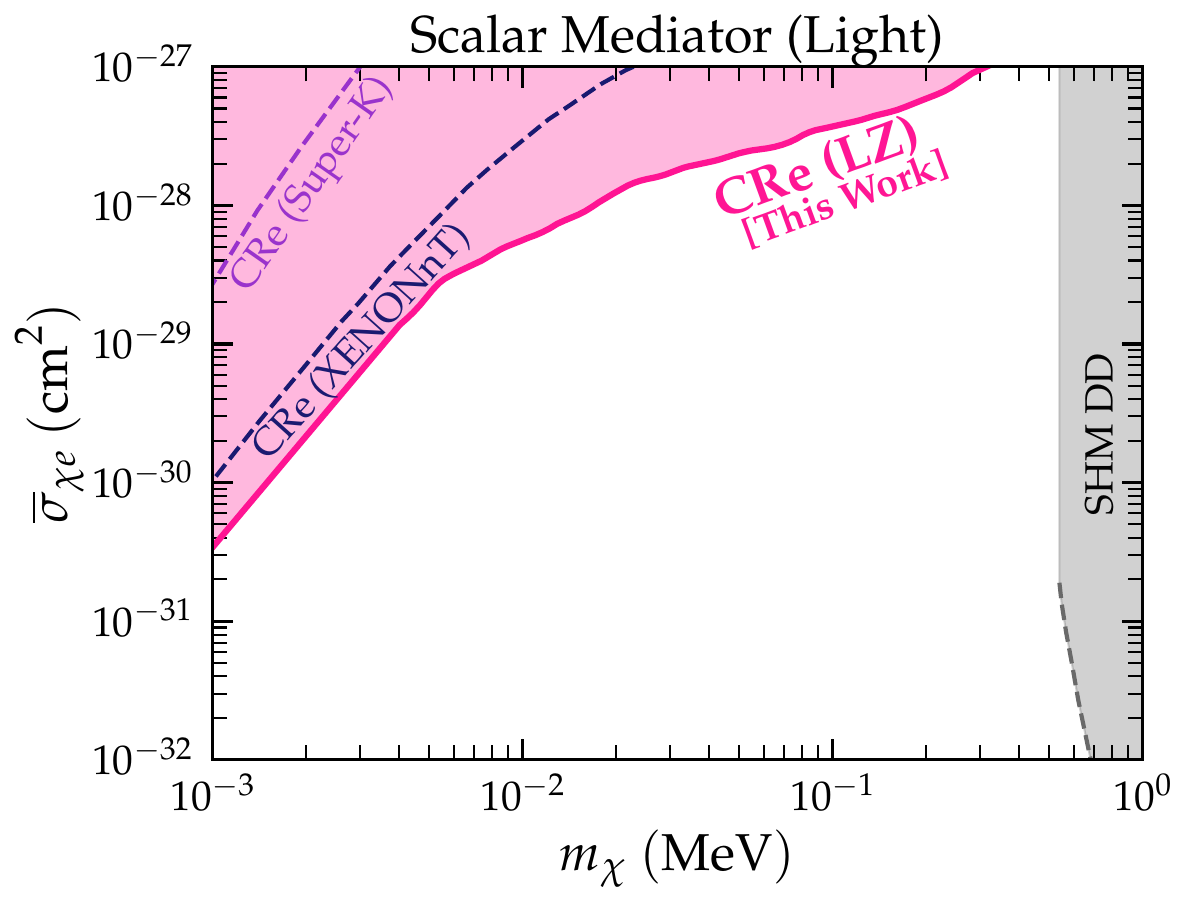}}
    \caption{$2\sigma$ C.L. constraints on CRe boosted DM in the light mediator limit ($F_{\rm DM}(q)=(\alpha m_e)^2/q^2$) with (a) {\bf vector} mediator and (b)  {\bf scalar} mediator from LZ 2025 are shown by the magenta shaded region. Existing constraints from the BDM search at XENONnT \cite{Guha:2024mjr} (dashed blue line), Super-k \cite{Bardhan:2022bdg} (dashed purple line), utilizing plasmon enhancement in SENSEI \cite{Liang:2024xcx}(dark cyan dashed lines), and the DAMIC 2025 \cite{DAMIC-M:2025luv} limit for Galactic halo DM (gray shaded region) are also portrayed. We also display astrophysical constraints like supernova cooling \cite{Chang:2018rso} (orange shaded region) and red giant \cite{Hardy:2016kme} (dark red solid line).}
    \label{fig:LZ_light}
\end{figure*}

The constraints on CRe boosted DM change drastically while considering energy dependent cross sections. 
To investigate these we consider the earlier mentioned two scenarios: vector and scalar mediated interaction between DM and $e$.
While analyzing the energy dependent cross sections the mass scale of the associated mediator is crucial in deciding the BDM flux and the recoil spectra.
We first display our $2\sigma$ C.L. constraints in the heavy mediator limit i.e. $F_{\rm DM} (q)=1$ obtained from LZ 2025 in Fig.~\ref{fig:LZ_heavy} for (a) vector mediator ({\bf left panel})  and (b) scalar mediator ({\bf right panel}).
To signify our limits and existing constraints we follow the same color convention as discussed earlier.
The combined
DD limits  \cite{SuperCDMS:2018mne,Crisler:2018gci,Essig:2012yx} for Galactic halo DM (gray shaded region) are also shown for both of the aforementioned plots.
For the vector mediator scenario, our obtained limit is stronger than the earlier bound from XENONnT \cite{Guha:2024mjr} (dashed blue line) by almost a factor of $\sim 2$.
We also highlight the constraints on CRe boosted DM from PandaX-4T \cite{PandaX:2024pme} and SENSEI \cite{Liang:2024xcx}(through plasmon enhancement) represented by green and dark cyan dashed lines respectively.
It is worth highlighting that, in the heavy mediator limit, LZ 2025 constraints with energy dependent cross sections are $\gtrsim \mathcal{O}(10^2)$ times stronger than the same with a constant cross section in the low ($\sim$keV) DM mass region.
On the other hand, the LZ limit obtained with a scalar mediator ({\bf right panel}) is weaker than that with vector mediator ({\bf left panel}) in low mass ($m_\chi\lesssim 10^{-1}$ MeV) range. 
This can be perceived from the associated $E_R$ suppression (see Eq.~\eqref{eq:sig_scalar}) in the differential cross section for scalar mediator \cite{Bardhan:2022bdg}. 
Similar to the case of constant cross section, here also we find an attenuation ceiling on the constraint region for higher values of $\bar{\sigma}_{\chi e}$. The differences in the shape of the attenuation ceiling for vector and scalar interactions, compared to the constant one, can also be attributed to the strong dependence of the DM interaction rate on the energies of the involved particles.
For both scalar and vector interaction, Super Kamiokande (Super-k) \cite{Bardhan:2022bdg} places stronger limits
on BDM interaction in the heavy mediator limit.
Since the upper limits from Super-K  fall below the lowest $\bar{\sigma}_{\chi e}$ values covered by the aforementioned plots, they are not shown explicitly to maintain the clarity of the figures.

On the other hand, $2\sigma$ C.L. constraints from  LZ 2025 in the light mediator limit i.e. $F_{\rm DM} (q)=(\alpha m_e)^2/q^2$ are shown in Fig.~\ref{fig:LZ_light} for the (a) vector mediator ({\bf left panel})  and (b) scalar mediator ({\bf right panel}).
Again, we follow the same color convention to represent the obtained limits and the existing constraints.
Note that, for this scenario, the obtained limits are significantly weaker than those obtained in the heavy mediator limit, due to the suppression in the differential cross section from $F_{\rm DM}$.
This leads to a significantly smaller amount of incoming BDM flux for $m_\chi\gtrsim 1$ MeV \cite{Bardhan:2022bdg} and consequently a suppressed event rate.
Note that our obtained limits in this case are also stronger than the previously obtained BDM constraints from XENONnT \cite{Guha:2024mjr, Bardhan:2022bdg} (dashed blue line), and Super-k \cite{Bardhan:2022bdg} (dashed purple line).
However, our limit is weaker than the BDM constraint from SENSEI \cite{Liang:2024xcx} (dark cyan dashed lines), which benefits from plasmon enhancement.

For the light mediator case, even with a larger cross section ($\bar{\sigma}_{\chi e}\gtrsim 10^{-26}~{\rm cm^2}$), the attenuation effect is found to be negligible. Since we do not observe any attenuation ceiling for this case in our region of interest, we refrain from showing it in the constraint plot.

A complementary bound from solar reflected DM is also applicable for our scenario and is expected to exclude $\bar{\sigma}_{\chi e}\lesssim 10^{-31}~{\rm cm}^2$ and $1~{\rm keV}\lesssim m_\chi\lesssim10$ MeV \cite{An:2017ojc,Emken:2024nox}. 
However, in the absence of any rigorous analysis of purely electrophilic solar reflected DM with a vector or scalar mediator in the existing literature, we refrain from showcasing the solar reflected DM limit.
For the ultralight mediator case, light mediators are constrained from cooling of astrophysical bodies like the Sun, red giant (RG), and supernova \cite{Hardy:2016kme,Tsai:2022jnv}.
The recasted limit on the $m_\chi$ vs $\bar{\sigma}_{\chi e}$ plane, from RG cooling (dark red solid line) excludes $m_\chi\lesssim 0.1$ MeV for the vector mediator.
However, in the scalar mediator scenario, the same limit is expected to exclude $\bar{\sigma}_{\chi e}\gtrsim10^{-42}{\rm cm}^2$ for our mass range of interest  \cite{Hardy:2016kme, Knapen:2017xzo}.
Hence, we do not explicitly showcase the shaded region corresponding to this limit for the light scalar mediator case to avoid overcrowding the plot.
On the other hand, supernova cooling (orange shaded regions) constrains $10^{-32}{\rm cm}^2 \lesssim \bar{\sigma}_{\chi e}\lesssim 10^{-38}{\rm cm}^2$ for $m_\chi=1$ keV for the ultralight vector mediator \cite{Knapen:2017xzo,Bardhan:2022bdg}.
Cosmological constraints from big bang nucleosynthesis \cite{Krnjaic:2019dzc} exclude fermion DM with $m_\chi\lesssim 8$ MeV, although they depend on the thermal history and can be surpassed in an extended DM sector.
{Our obtained limit from LZ 2025 can serve as a complementary direct probe of them for the light vector and scalar mediated DM.}

As discussed before, Super-k \cite{Bardhan:2022bdg} and IceCube \cite{Cappiello:2024acu} so far set the most stringent constraints on CRe boosted (sub-) MeV DM in the heavy mediator limit, thanks to their humongous target size.
For example, for the constant cross section $\bar{\sigma}_{\chi e}\gtrsim 10^{-32}~{\rm cm^2}$ is excluded from Super-k for $m_\chi \sim 1$ MeV \cite{Ema:2018bih}.
For the vector (scalar) mediated cross section  Super-k excludes $\bar{\sigma}_{\chi e}\gtrsim 10^{-34}~{\rm cm^2}$ ($\bar{\sigma}_{\chi e}\gtrsim 10^{-33}~{\rm cm^2}$) for $m_\chi \sim 1$ MeV in the heavy mediator limit \cite{Bardhan:2022bdg}.
However, in the light mediator limit, neutrino detector constraints are significantly weakened.
For vector (scalar) mediated cross section  Super-k excludes $\bar{\sigma}_{\chi e}\gtrsim 10^{-29}~{\rm cm^2}$ ($\bar{\sigma}_{\chi e}\gtrsim 10^{-28}~{\rm cm^2}$) for $m_\chi \sim 1$ keV in the light mediator limit \cite{Bardhan:2022bdg}.
The threshold of Super K (IceCube) is $\mathcal{O}(100)$ MeV ($\mathcal{O}(500)$ GeV) which leads to a suppressed ($\sim 1/q^2$) recoil rate  in the light mediator limit. 
On the other hand, LZ benefits from its very low threshold energy ($\sim 1$ keV) and can place a stronger constraint in the light mediator limit. 
Though the same parameter space is strongly constrained by stellar-cooling limits and novel table-top experiments such as SENSEI, our results demonstrate the complementary sensitivity of LZ 2025 to CRe-boosted dark matter in the light-mediator regime.

\begin{figure}[!tbh]
    \centering
    \includegraphics[scale=0.45]{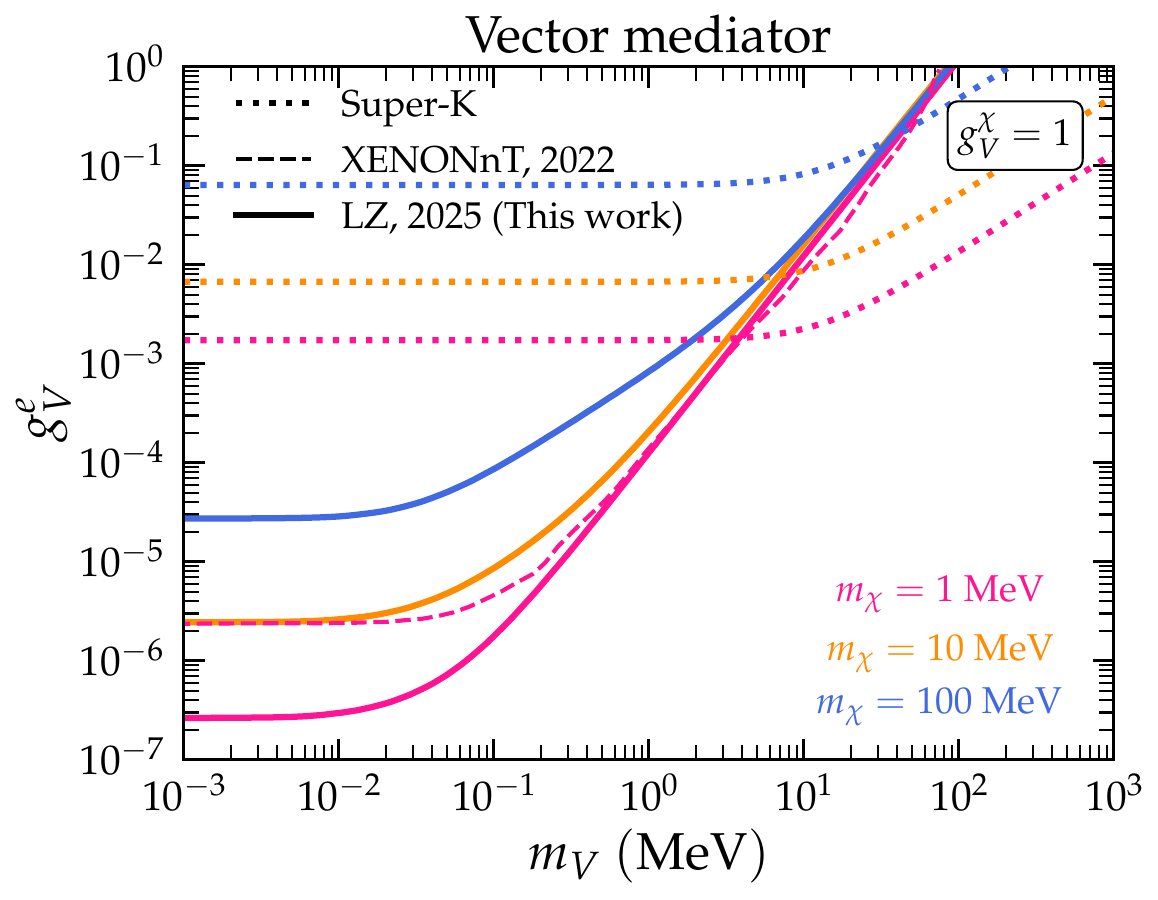}
    \caption{$2\sigma$ C.L. constraints in the {\bf vector} mediator mass vs coupling parameter space for CRe boosted DM from LZ 2025. Limits corresponding to $m_\chi=1$ MeV,~10 MeV, and $100$ MeV are represented by the region above the magenta, orange and blue solid lines respectively.}
    \label{fig:med_vec}
\end{figure}
Owing to the huge differences in different detector thresholds, it is also crucial to properly identify the mediator mass range while comparing the limits from DM and neutrino detectors. For example, ``heavy mediator" limit in Super-k signifies the mediator mass to be $\gtrsim \mathcal{O}(100)$ MeV, whereas the same limit in the LZ can be realized when the mediator mass is $\gtrsim \mathcal{O}(100)$ keV.
Thus, for concrete analysis, it is essential to consider minimal benchmark DM models featuring the earlier mentioned mediators.
As an example, we choose a minimal model with fermion DM $\chi$ and a beyond the standard model (BSM) vector mediator $V$ and write the following Lagrangian,
\begin{eqnarray}
    \mathcal{L}_{\rm int} \supset -g^\chi_V \bar{\chi} \gamma^\mu \chi V_\mu - g^e_V \bar{e} \gamma^\mu e V_\mu\,,
\end{eqnarray}
where$g^e_V~(g^\chi_V)$ is the respective coupling of $V$ with electron (DM). The mediator mass is denoted as $m_V$. The reference NR cross section is then given by,

\begin{eqnarray}
    \bar{\sigma}_{\chi e} =\frac{4(g^e_V g^\chi_V)^2 \mu_{\chi e}^2}{\pi (q_{\rm ref}^2+m_V^2)^2}
\end{eqnarray}
In Fig.~\ref{fig:med_vec} we present our $2\sigma$ C.L. constraints in the vector mediator parameter space ($m_V$ vs $g^e_V$ plane) for CRe boosted DM from LZ 2025.
Here we fix the DM coupling $g^\chi_V$ to be 1.
We display our exclusion limits for $m_\chi=1$ MeV,~10 MeV, and $100$ MeV depicted by the regions above the magenta, orange, and blue solid lines respectively.
For a quick comparison, our obtained limit on $g^e_V$ for $m_\chi=1$ MeV at $m_V\lesssim 10^{-2}$ MeV is stronger than XENONnT \cite{Guha:2024mjr} by a factor of $\sim 6$.
In the same plane we also showcase the Super-k limit on $g^e_V$ depicted by dotted lines (see.Appendix \ref{sec:supk} for Super-k analysis).
Note that, Super-k limits for $m_V\lesssim 1$ MeV are $\mathcal{O}(2)$ weaker  than our limit.
A similar analysis with a scalar mediator can also be performed (see Appendix \ref{sec:scal_med}).

For the light mediators discussed in the context of Fig.~\ref{fig:med_vec}, there also exist constraints from 
BaBar ($g_{V}^e\gtrsim10^{-4}$) \cite{BaBar:2014zli,Bauer:2018onh}, invisible decay from NA64 \cite{Banerjee:2019pds}
and beam dump experiments \cite{Bauer:2018onh}.
The limits from the collider and fixed-target experiments depend strongly on the ultraviolet (UV) completion of the benchmark model and change drastically depending on the DM mass and other interactions of the mediator.
For example,  the invisible decay constraints coming from BaBar \cite{BaBar:2014zli,Bauer:2018onh}, and NA64 invisible \cite{Banerjee:2019pds}
are relevant as long as the mediator mass $<2 m_\chi$, so that the mediator
can decay dominantly to the dark sector.
Thus, the parameter space does not encounter any single constraint contour from these constraints, rather a different set of benchmark-dependent constraints are applicable.
BaBar ($g_{V}^e\gtrsim10^{-4}$) \cite{BaBar:2014zli,Bauer:2018onh} is expected to exclude the parameter space $m_V\sim 2$MeV-$10$GeV ($m_V\sim 200$MeV-$10$GeV) for $m_\chi=1$ MeV (100 MeV).
On the other hand, the visible decay constraints (beam
dump \cite{Bauer:2018onh}) are derived assuming that the mediator decays to SM particles with $\sim 100\%$ branching ratio and constrains the light mediator in the range $1-100$ MeV. 
Again, for our chosen set of DM mass, the beam dump constraint applies only for $m_\chi =100$ MeV.
Hence, we refrain from recasting these benchmark-dependent limits in our parameter space to keep the figure uncluttered. 
Apart from these stellar cooling constraint excludes $m_V< 0.4$ MeV \cite{Hardy:2016kme}, although the constraint reported in the existing literature is derived without considering mediators coupled with the dark sector.
However, these constraints are often associated with astrophysical uncertainties and are dependent on stellar modeling.
Thus our obtained limits are complementary to these limits and, in some parameter space (for $m_\chi \lesssim 10$ MeV, and $m_V\lesssim 1$ MeV), stronger than the existing limits.
Thus, LZ 2025 data can be extremely helpful for understanding the particle nature of an electrophilic light DM accelerated by the cosmic rays and help to address one of the key puzzles of the Universe.

{\bf Conclusion:}
CRe upscattering of light DM makes them detectable in the underground detectors which  can probe the previously inaccessible DM  parameter space.
In this work we provide the updated constraint on such electrophilic DM analyzing the latest LZ data \cite{LZ:2025zpw}.
We show that LZ places a constraint on the DM-e cross section stronger than the previous DD limit from  XENONnT \cite{Guha:2024mjr} .
For such boosted DM the energy dependencies in their respective cross section have a pivotal role in the obtained rates and consequently in the constraints from DD experiments.
To scrutinize this issue we also consider realistic cross sections with minimal DM models featuring vector and scalar mediators.
Through exhaustive numerical scans within such frameworks, we present the limits obtained from LZ 2025 on CRe boosted DM. 
We also incorporate the Earth's attenuation effect on BDM flux and implement a full numerical treatment to obtain constraints.
We observe that, with the realistic cross sections, the obtained limits strongly depend on the mediator mass, 
compelling us to rigorously analyze it with realistic models featuring BSM mediators.
Even in the light-mediator regime, LZ provides a complementary probe of the parameter space alongside constraints from other experiments.
In the future, experiments such as DARWIN, JUNO, DUNE, and Hyper-Kamiokande will frame the pathway of further investigation of the low mass boosted DM parameter space.

\section*{Acknowledgement}
We acknowledge the
use of the high-performance computing facility, the Bhaskara Cluster, at the Department of Physics, IISER Bhopal,
which significantly facilitated the completion of this research.
SJ is supported
by the National Natural Science Foundation of China (12425506, 12375101, 12090060, and
12090064) and the SJTU Double First Class start-up fund (WF220442604). 
AM acknowledges financial support from the Government of India through the Prime Minister Research Fellowship (PMRF) scheme (ID No.: 0401970).
\appendix
\counterwithin{figure}{section}


\section{Attenuated BDM flux}
\label{sec:attenuation}

\begin{figure}[!h]
    \centering
    \subfigure[\label{a1}]{
    \includegraphics[scale=0.6]{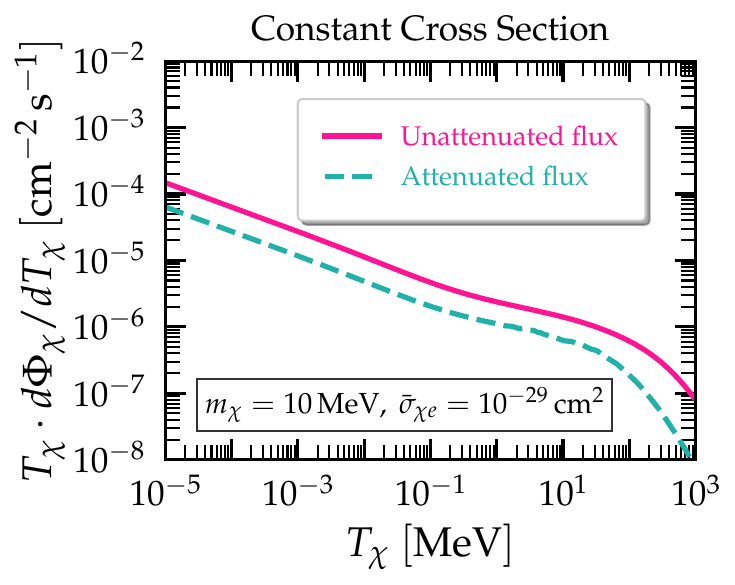}}
    \subfigure[\label{a2}]{
    \includegraphics[scale=0.6]{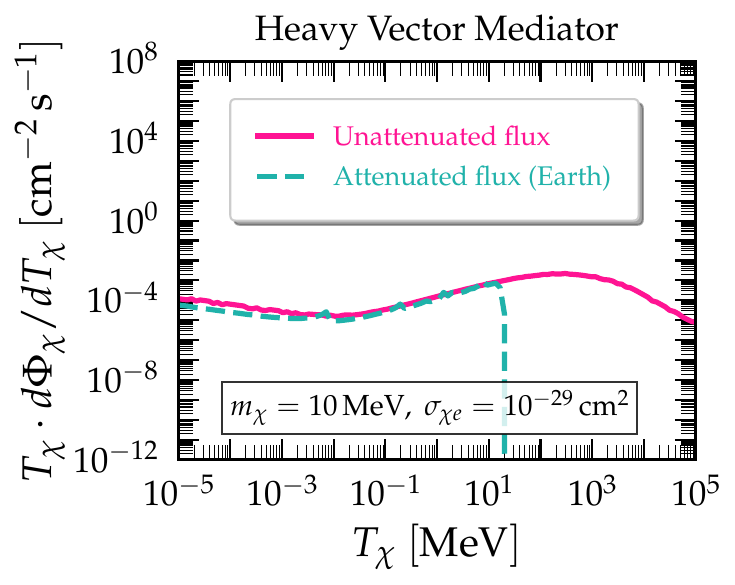}}
    \subfigure[\label{a3}]{
    \includegraphics[scale=0.6]{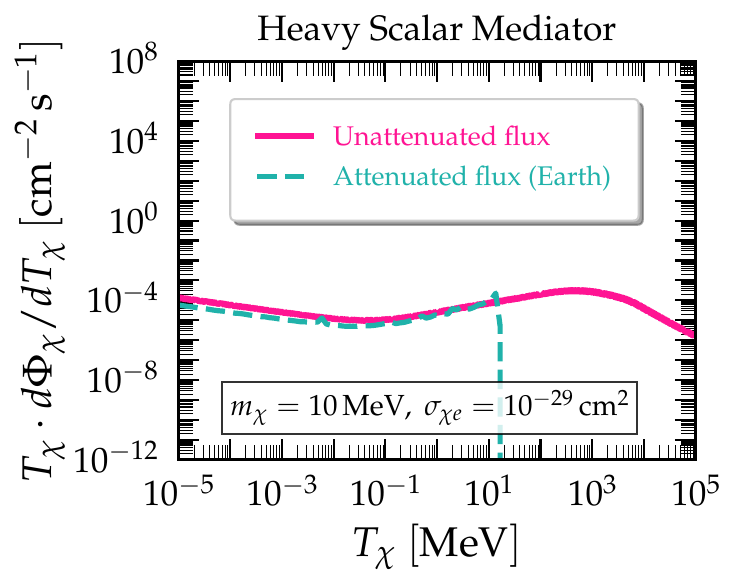}}
    \caption{{Comparison between incoming BDM fluxes at  LZ detector \cite{LZ:2025zpw} with and without the attenuation effect depicted by green dashed and red solid lines, respectively for (a) constant, (b) heavy vector mediated, and (c) heavy scalar mediated interactions. We choose a benchmark $m_\chi=10$ MeV and $\bar{\sigma}_{\chi e}=10^{-29}{\rm cm}^2$.}}
    \label{fig:flux_heavy}
\end{figure}

\begin{figure}[!h]
    \centering
    \subfigure[\label{a4}]{
    \includegraphics[scale=0.6]{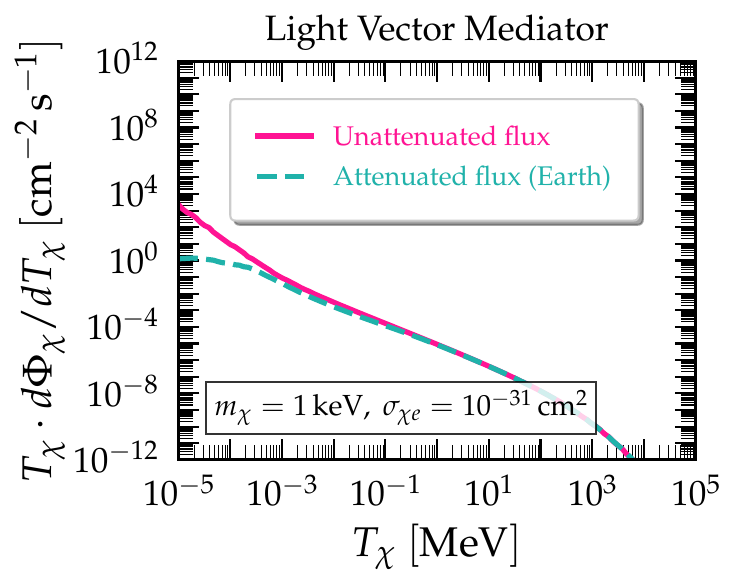}}
    \subfigure[\label{a5}]{
    \includegraphics[scale=0.6]{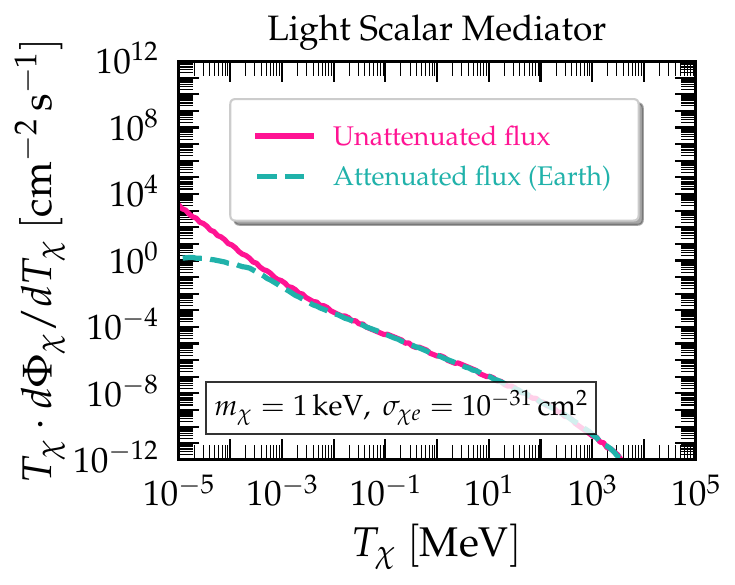}}
    \caption{{ Comparison between incoming BDM fluxes at  LZ detector \cite{LZ:2025zpw} with and without the attenuation effect depicted by green dashed and red solid lines, respectively for (a) light vector mediated and (b) light scalar mediated interactions. We choose a benchmark $m_\chi=1$ keV and $\bar{\sigma}_{\chi e}=10^{-31}{\rm cm}^2$.}}
    \label{fig:flux_light}
\end{figure}

Following the steps mentioned in the main text we obtain the attenuated flux of CRe boosted DM at the detector. 
In Fig.~\ref{fig:flux_heavy} we show the incoming BDM flux with and without the attenuation effect depicted by green dashed and red solid lines, respectively for $m_\chi=10$ MeV and $\bar{\sigma}_{\chi e}=10^{-29}{\rm cm}^2$.
The fluxes corresponding to constant, heavy vector mediated and heavy scalar mediated interactions are displayed in 
Fig.~\ref{a1}, Fig.~\ref{a2}, and Fig.~\ref{a3}, respectively.
Similarly, in Fig.~\ref{fig:flux_light} we compare the incoming BDM flux with and without the attenuation effect for $m_\chi=1$ keV and $\bar{\sigma}_{\chi e}=10^{-31}{\rm cm}^2$ for light vector mediated [Fig.~\ref{a4}] and light scalar mediated [Fig.~\ref{a5}] interactions.

Note that, for the constant cross section the attenuated flux differs slightly from the unattenuated flux with a moderate cross section as the DM mean free path ($l\sim (n_e \bar{\sigma}_{\chi e})^{-1}$) becomes comparable with the Earth's radius.
For similar reasons, the attenuated fluxes for the case of heavy vector and heavy scalar mediated interactions differ significantly from the unattenuated flux at higher $T_\chi$.
With higher values of $\bar{\sigma}_{\chi e}$, the fluxes decrease more and lead to a suppression in the event rate. For sufficiently large cross sections, the incoming DM particles lose a substantial fraction of their kinetic energy while propagating through the Earth and consequently fail to generate observable recoil signals at the detector, resulting in the upper ceilings observed in the constraint plots.
Needless to say, with smaller values of $\bar{\sigma}_{\chi e}$, the effect of attenuation becomes negligible and hence the upper limit of allowed DM cross sections remains almost similar to that obtained without considering the attenuation effect.  

On the other hand, for the light mediator case, the attenuated flux does not exhibit any significant difference from the unattenuated flux due to the 
momentum suppressed interaction rate with Earth's electrons. 
Hence, for this case also, the upper limit of the allowed DM cross sections coincides with that obtained without considering the attenuation effect.
Even varying $\bar{\sigma}_{\chi e}$ up to  $10^{-25}{\rm cm}^2$ we do not find any corresponding attenuation ceiling
and hence do not include the attenuation ceiling in Fig.~\ref{fig:LZ_light}.

\section{Differential cross section and recoil spectra}
\label{sec:recoil}

\begin{figure}[htp]
    \centering
    \subfigure[\label{h}]{
    \includegraphics[scale=0.35]{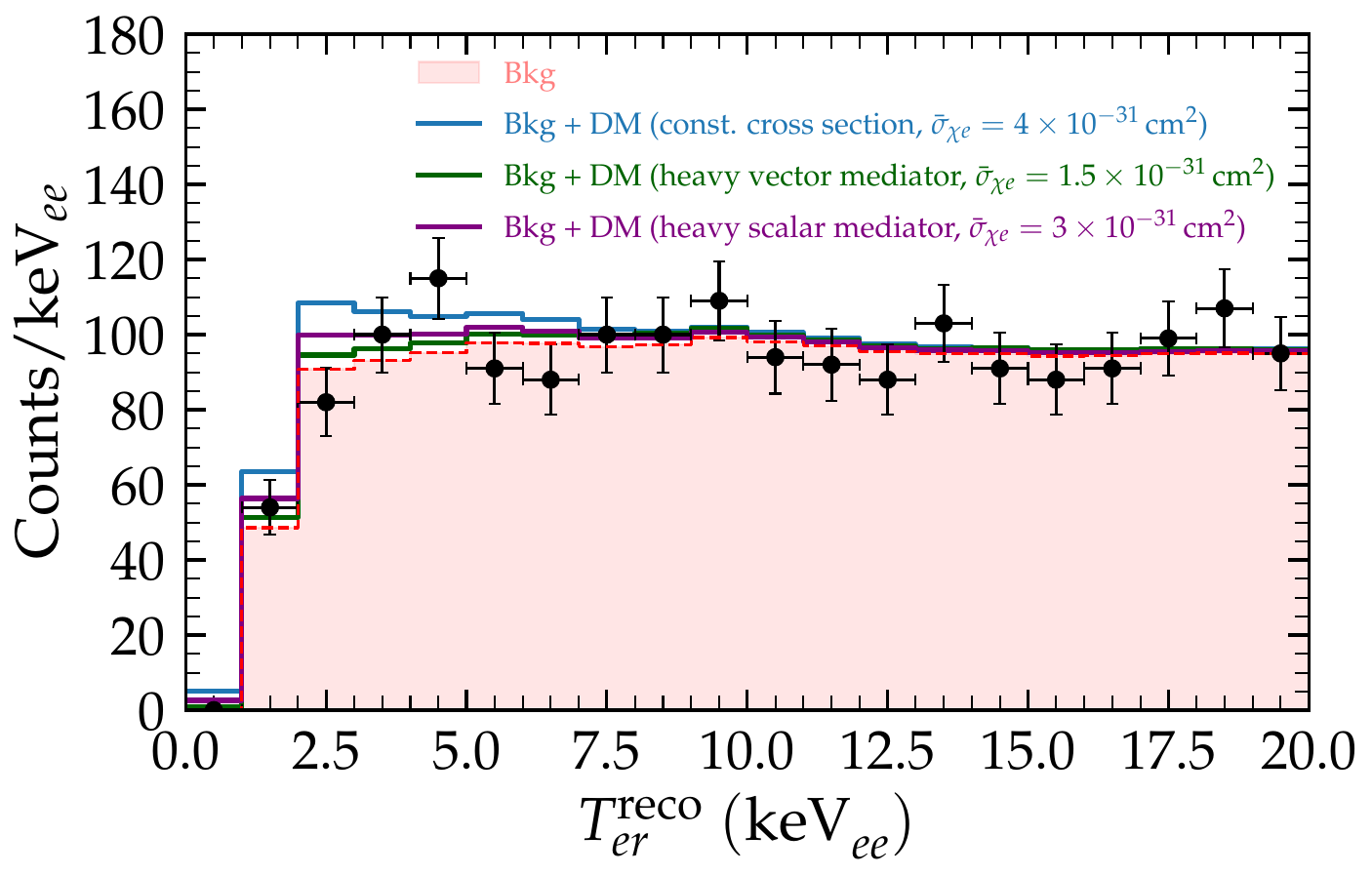}}
    \subfigure[\label{l}]{
    \includegraphics[scale=0.35]{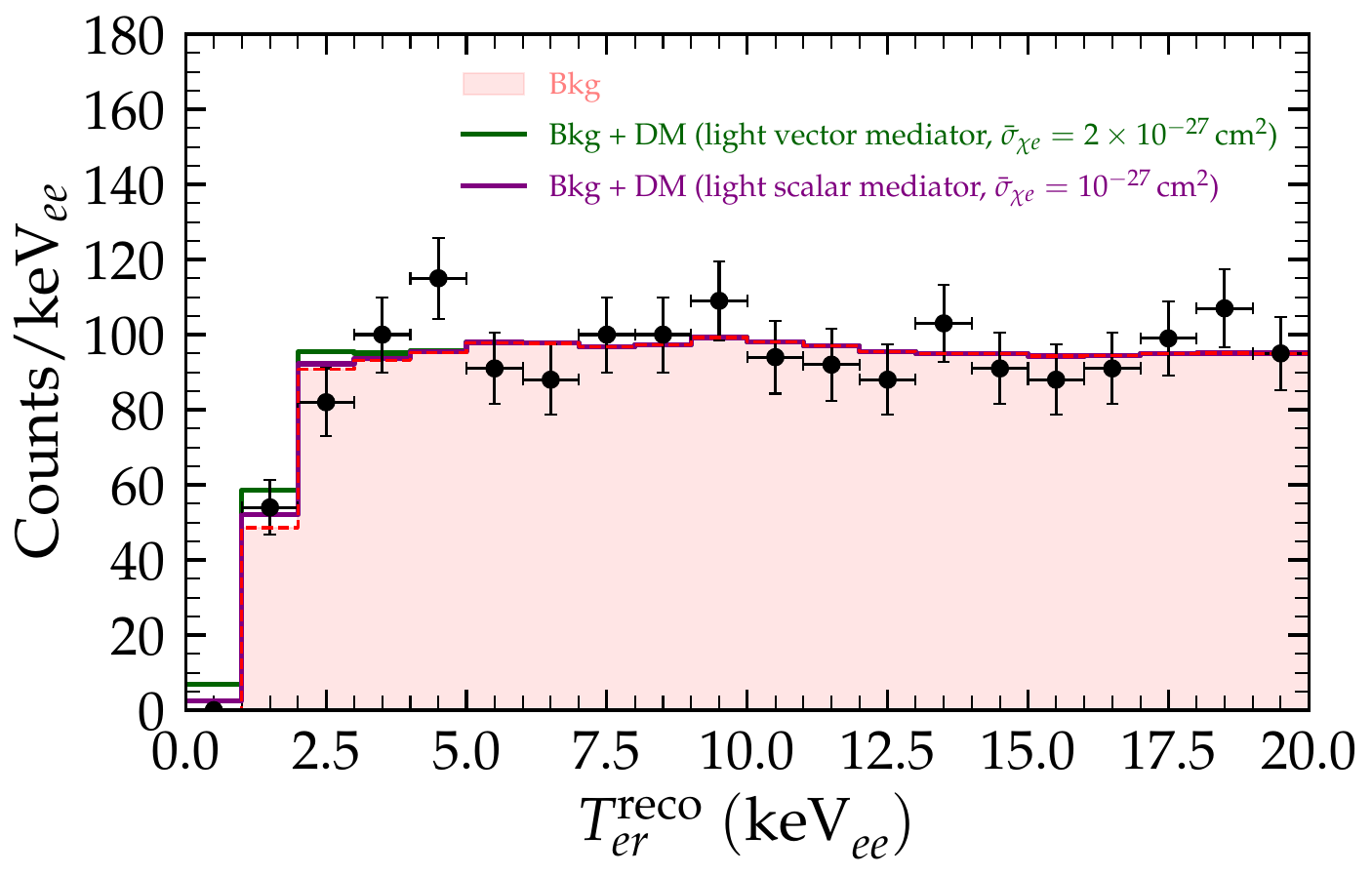}}
    \caption{Expected recoil signature of CRe boosted DM in the LZ detector \cite{LZ:2025zpw} considering the attenuation effect for the (a) heavy mediator with $m_\chi=1$ MeV, and (b) light mediator with $m_\chi=0.1$ MeV.}
    \label{fig:recoil_spectra}
\end{figure}

The differential cross section for DM scattering with target electron in the detector is given as,
\begin{widetext}
    \begin{eqnarray}
     \left(\frac{d\sigma_{\chi e}}{dT_{{er}}}\right)_{\rm s} &=& \bar{\sigma}_{\chi e} \left (\dfrac{m_e}{4\mu_{e \chi}^2} \right ) \dfrac{(2m_e+T_{er})(2 m_\chi^2+m_e T_{er})}{s_{\chi}T_{er}^{\rm max}},\label{eq:sig_scalar}\\ 
    \left(\frac{d\sigma_{\chi e}}{dT_{er}}\right)_{\rm v} &=& \bar{\sigma}_{\chi e} \left (\dfrac{m_e}{2\mu_{e \chi}^2}\right ) \dfrac{2 m_e (m_\chi+T_\chi)^2 -T_{er} ((m_e+m_\chi)^2+2 m_e T_\chi )+m_e T_{er}^2 }{s_{\chi}T_{er}^{\rm max} },\label{eq:sig_vector}
\end{eqnarray}
\end{widetext}
for the scalar-mediated and vector-mediated cross section, respectively.
For this analysis, we use the following smearing function 
\begin{eqnarray}
    \mathcal{G}(T_{er},T_{er}^{\rm reco}) =\frac{1}{\sqrt{2\pi}\sigma}\exp{\left(-\dfrac{(T_{er} -T_{er}^{\rm reco})^2}{2\sigma^2} \right)},
\end{eqnarray}
$\sigma$ being the resolution power mentioned earlier \cite{Pereira:2023rte}.

The expected recoil signature is shown in Fig.~\ref{fig:recoil_spectra}. The red shaded region and the black dots with error bars signify the background and observed events reported by LZ 2025 \cite{LZ:2025zpw}, respectively.
In Fig.~\ref{h} we show the event spectra for different cross section types with a fixed mass $m_\chi=1$ MeV depicted by different colored lines. 
The chosen cross sections are $4\times 10^{-31}~{\rm cm}^2$, $1.5\times 10^{-31}~{\rm cm}^2$, and $3\times 10^{-31}~{\rm cm}^2$ for constant (steel blue), vector (green), and scalar (purple) mediated cross sections, respectively.
On the other hand, the recoil spectra with a light mediator limit is displayed 
in Fig.~\ref{l}.
Here we consider $m_\chi=0.1$ MeV and $\bar{\sigma}_{\chi e}=2\times 10^{-27}~{\rm cm}^2$, and $10^{-27}~{\rm cm}^2$ for vector (green) and scalar (purple) mediated cross sections, respectively.
It is worth noting the role of the interaction type and the mediator mass role in deciding the  recoil signature in the aforementioned plots.


\section{Statistical Analysis}
\label{sec:chi}

The statistical interpretation of the LZ WS2024 data is performed using a binned spectral analysis based on a Poisson-likelihood $\chi^2$ framework~\cite{Almeida:1999ie}. The test statistic is defined as
\begin{align}
\chi^2(\vec{\mathcal{S}};\gamma_k) &= 2 \sum_i \Bigg[
R_\mathrm{pred}^i(\vec{\mathcal{S}};\gamma_k) - R_\mathrm{exp}^i \notag \\
&+ R_\mathrm{exp}^i \ln\!\bigg(
\frac{R_\mathrm{exp}^i}{R_\mathrm{pred}^i(\vec{\mathcal{S}};\gamma_k)}
\bigg)
\Bigg]  + \sum_k \left( \frac{\gamma_k}{\sigma_{\gamma_k}} \right)^2 ,
\end{align}
where $R_\mathrm{exp}^i$ denotes the observed number of events in the $i$-th recoil-energy bin, taken from Fig.~2 of Ref.~\cite{LZ:2025zpw}. The predicted event rate in each bin is obtained by combining the theoretically simulated dark matter signal ($R_{\chi e}^i(\vec{\mathcal{S}})$) with the individual background components as reported by the LZ Collaboration~\cite{LZ:2025zpw}, and is expressed as,
\begin{widetext}
\begin{align}
R_\mathrm{pred}^i(\vec{\mathcal{S}};\gamma_k) &= R_{\chi e}^i(\vec{\mathcal{S}}) + (1+\gamma_{^{214}\mathrm{Pb}})\, R_{^{214}\mathrm{Pb~}\beta}^i\notag  + (1+\gamma_{\mathrm{Kr/Ar}})\, R_{^{85}\mathrm{Kr}+^{39}\mathrm{Ar}+\beta_s+\gamma}^i  + (1+\gamma_{\nu\mathrm{ER}})\, R_{\nu_\odot\mathrm{ER}}^i \notag \\
&\quad+ (1+\gamma_{2\nu\beta\beta})\, R_{^{136}\mathrm{Xe}\,2\nu\beta\beta}^i \notag  + (1+\gamma_{\mathrm{Po/Pb}})\, R_{^{218}\mathrm{Po}+^{212}\mathrm{Pb~}\beta_s}^i + (1+\gamma_{\mathrm{H/C}})\, R_{^{3}\mathrm{H}+^{14}\mathrm{C}}^i   \notag \\
&\quad+  (1+\gamma_{\mathrm{DEC}})\, R_{^{124}\mathrm{Xe}\,\mathrm{DEC}}^i + (1+\gamma_{\mathrm{EC}})\, R_{^{127}\mathrm{Xe}+^{125}\mathrm{Xe~EC}}^i .
\end{align}
\end{widetext}

The nuisance parameters $\gamma_k$ encode uncertainties in the normalization of individual background components and are constrained by Gaussian priors with widths~\cite{LZ:2025zpw}

\begin{align}
\sigma_{^{214}\mathrm{Pb}} &= \frac{130}{1099}, \qquad
\sigma_{\mathrm{Kr/Ar}} = \frac{33}{237}, \qquad
\sigma_{\nu\mathrm{ER}} = \frac{9}{151}, \notag \\
\sigma_{2\nu\beta\beta} &= \frac{16}{107}, \qquad
\sigma_{\mathrm{Po/Pb}} = \frac{11.0}{92.1}, \qquad
\sigma_{\mathrm{H/C}} = \frac{3.4}{61.2}, \notag \\
\sigma_{\mathrm{DEC}} &= \frac{4.0}{20.0}, \qquad
\sigma_{\mathrm{EC}} = \frac{0.7}{3.2}.
\end{align}

These nuisance parameters are marginalized over in the $\chi^2$ minimization procedure for each choice of the new-physics parameter set $\vec{\mathcal{S}}$ i.e. each set of $\{m_\chi, \bar{\sigma}_{\chi e}\}$ values.

\section{DM with light scalar mediators}
\label{sec:scal_med}

%
\begin{figure}[!tbh]
    \centering
    \includegraphics[scale=0.45]{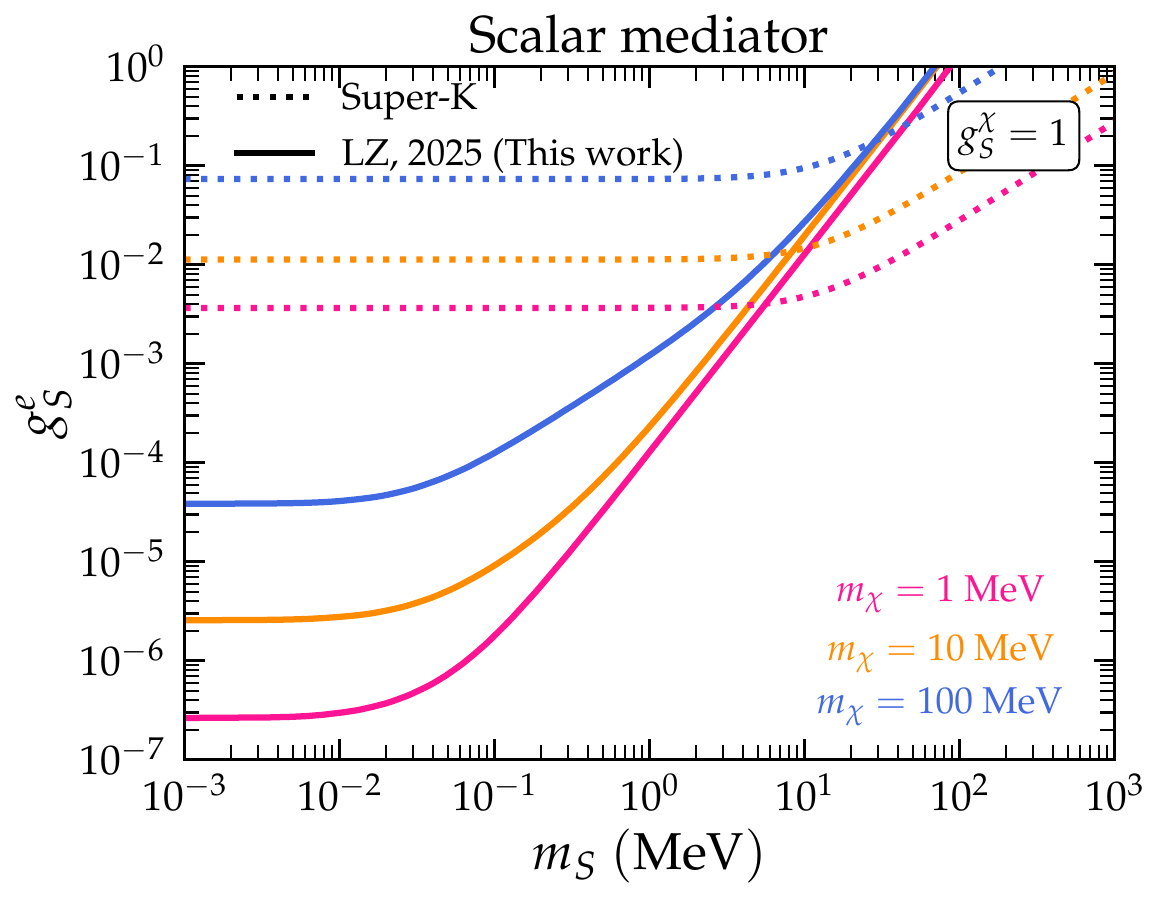}
    \caption{$2\sigma$ C.L. constraints in the {\bf scalar} mediator mass vs coupling parameter space for CRe boosted DM from LZ 2025. Limits corresponding to $m_\chi=1$ MeV,~10 MeV, and $100$ MeV are represented by the region above the magenta, orange and blue solid lines respectively.}
    \label{fig:med_scal}
\end{figure}
Here, we choose a minimal model with fermion DM $\chi$ and a scalar mediator $S$ with the following Yukawa like Lagrangian,
\begin{eqnarray}
    \mathcal{L}_{\rm int} \supset -g^\chi_S S \bar{\chi} \chi  - g^e_S S\bar{e} e ,
\end{eqnarray}
where $g^e_S~(g^\chi_S)$ is the respective coupling of $S$ with electron (DM). The mediator mass is denoted as $m_S$. The reference NR cross section is then given by,
\begin{eqnarray}
    \bar{\sigma}_{\chi e} =\frac{(g^e_S g^\chi_S)^2 \mu_{\chi e}^2}{\pi (q_{\rm ref}^2+m_S^2)^2}
\end{eqnarray}
The corresponding $2\sigma$ C.L. constraints in the scalar mediator parameter space ($m_S$ vs $g^e_S$ plane) for CRe boosted DM from LZ 2025 is presented in Fig.~\ref{fig:med_scal}.
Here, we also fix the DM coupling $g^\chi_S$ to be 1 and showcase our exclusion limits for $m_\chi=1$ MeV,~10 MeV, and $100$ MeV depicted by the regions above the magenta, orange and blue solid lines, respectively.

\section{Super-k constraints}
\label{sec:supk}
Super-K constraints for this work are obtained following the methodology prescribed in Ref.~\cite{Ema:2018bih}.
Super-K is a 50 kton water Cherenkov detector   \cite{Super-Kamiokande:2017dch} and reported 4042 events observed in the first bin with $0.1~{\rm GeV}< E_R<1.33$~GeV with an efficiency $\mathcal{\epsilon}_{\rm SK}=0.93$ analyzing 2628.1 days of its data taken with a 161.9 kton-year exposure.
To obtain a conservative limit, we consider the predicted event in the energy range $N_{\rm SK}>4042$: 
\begin{eqnarray}
   n_T~ t_{\rm run} ~\mathcal{\epsilon}_{\rm SK}\times \int_{0.1{~\rm GeV}}^{1.33{\rm ~GeV}} \frac{d R}{d E_R} d E_R > 4042,
\end{eqnarray}
where $n_T~ t_{\rm run}=7.5\times 10^{34}\times$year represents total exposure \cite{Ema:2018bih}. The differential cross section is considered to be the vector (scalar) mediated one to obtain the limit on $g_V^e~(g_S^e)$ as shown in Fig.~\ref{fig:med_vec} (Fig.~\ref{fig:med_scal}).

\bibliography{ref}
\bibliographystyle{JHEP}

\end{document}